\documentclass[aps,prx,10pt, longbibliography, reprint,groupedaddress,superscriptaddress]{revtex4-2}

\usepackage[dvipsnames]{xcolor}
\usepackage{amssymb}
\usepackage{graphicx}
\usepackage{amsmath}
\usepackage{enumitem}
\usepackage{txfonts}
\usepackage[bookmarks=true,colorlinks,citecolor=blue,urlcolor=blue]{hyperref}
\usepackage{braket}
\usepackage{babel}
\usepackage{blindtext}

\usepackage[normalem]{ulem}
\newcommand{\cd}[1]{c^{\dagger}_{#1}}
\newcommand{\co}[1]{c^{}_{#1}}

\newcommand{\fd}[1]{f^{\dagger}_{#1}}
\newcommand{\fo}[1]{f^{}_{#1}}
\newcommand{\bd}[1]{b^{\dagger}_{#1}}
\newcommand{\bo}[1]{b^{}_{#1}}

\newcommand{\kvec}{\mathbf{k}}
\newcommand{\qvec}{\mathbf{q}}

\newcommand{\rvec}{\mathbf{r}}

\newcommand{\suS}[1]{\mathcal{S}_{#1}}

\begin{document}

\title{Fractionalization from Kinetic Frustration in Doped Two-Dimensional SU(4) Quantum Magnets}

\newcommand{\TUM}{\affiliation{Technical University of Munich, TUM School of Natural Sciences, Physics Department, 85748 Garching, Germany}}
\newcommand{\MCQST}{\affiliation{Munich Center for Quantum Science and Technology (MCQST), Schellingstr. 4, 80799 M{\"u}nchen, Germany}}
\newcommand{\ETH}{\affiliation{Institute for Theoretical Physics, ETH Zurich, Wolfgang-Pauli-Str. 27, 8093 Zurich, Switzerland}}

\author{Wilhelm Kadow} \TUM \MCQST
\author{Ivan Morera} \ETH
\author{Eugene Demler} \ETH
\author{Michael Knap} \TUM \MCQST
\date{\today}

\begin{abstract}
Separating electrons into emergent fractional quasiparticles is a hallmark of exotic quantum phases of matter with strong interactions. Understanding under which circumstances fractionalized excitations appear is a major conceptual challenge and can help realize long sought-after states, such as quantum spin liquids.
Here, we identify a distinct mechanism for fractionalization. Starting from the plaquette-ordered ground state of an SU(4) symmetric $t$-$J$ model at quarter filling on frustrated triangular lattices, we reveal a compelling interplay between order and fractionalization as a function of doping.
For hole doping, we find that the kinetic frustration can be relieved by fractionalizing the holes into fermionic spinons and bosonic holons: the holons minimize their kinetic energy when the spinons form a spinon Fermi surface. We support this mechanism analytically in the large-$N$ limit as well as numerically by simulating the SU(4) case with matrix product states on cylinder geometries and with variational Monte Carlo methods on system sizes up to $40\times40$. Conversely, electron doping drives the system into a ferromagnetic phase, akin to Nagaoka's theorem. We discuss possible experimental realizations in moiré heterostructures as well as ultracold atoms, and propose dynamical probes to search for key characteristics of the fractionalized quasiparticles.
\end{abstract}

\maketitle

\section{Introduction}

The interplay of strong interactions and frustration can give rise to fractionalized excitations where electronic or spin degrees of freedom reorganize into new emergent constituents, carrying fractional quantum numbers. Famous examples are quantum spin liquids~\cite{andersonResonatingValenceBonds1973, Savary2016, knolleFieldGuideSpin2019, broholmQuantumSpinLiquids2020}, which defy traditional ordering mechanisms but are dominated by quantum fluctuations instead. The discovery of exactly solvable models~\cite{kitaevFaulttolerantQuantumComputation2003, Kitaev06} and the development of symmetry classifications~\cite{wenQuantumOrdersSymmetric2002, wenQuantumFieldTheory2007} establish spin liquid phases as well-defined quantum phases of matter.
Spin liquids are expected in systems with strong interactions and geometric frustration, where the spins cannot simultaneously minimize all local constraints.
Rather than selecting a symmetry-breaking pattern, they remain quantum-disordered and are characterized by long-range entanglement and emergent gauge fields. An important distinction is between gapped spin liquids---for instance, when spinons occupy topological bands separated by a large gap~\cite{kalmeyerEquivalenceResonatingvalencebondFractional1987, wenChiralSpinStates1989, motrunichOrbitalMagneticField2006, gongGlobalPhaseDiagram2017, Kuhlenkamp2024, divic2024chiralspinliquidquantum}---and gapless ones, where fractionalized excitations persist down to the lowest energies. A conceptually simple gapless scenario is a spinon Fermi surface~\cite{wenQuantumOrdersSymmetric2002, motrunichVariationalStudyTriangular2005}.

Despite substantial progress, identifying microscopic Hamiltonians that robustly realize quantum spin liquids is a key challenge. In many candidate models, spin liquids appear only in narrow regions of parameter space and compete with nearby ordered phases, and their detection is complicated by the absence of a conventional local order parameter. In particular, although the spinon Fermi surface is conceptually simple at a mean-field level within the parton framework, a physically motivated spin model in which it is numerically established as the ground state remains sought after, underscoring the delicate role of gauge fluctuations beyond mean-field theory. These challenges motivate the search for new mechanisms to stabilize these fractionalized states.

\begin{figure}[b]
\centering
\includegraphics[trim={0cm 0cm 0cm 0cm},clip,width=\linewidth]{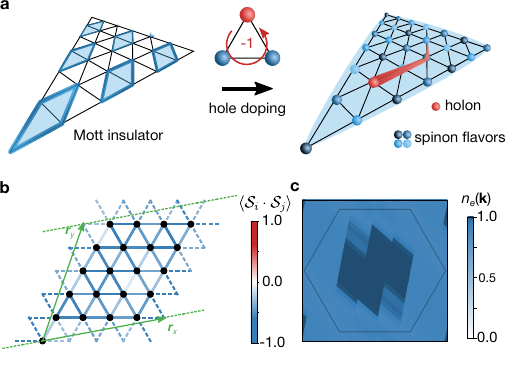}
\caption{\textbf{Kinetic frustration in the SU(4)-symmetric $t$-$J$ model.} \textbf{a)} In a Mott insulator with one electron per site, the SU(4)-symmetric model Eq.~\eqref{eq::Heisenberg_SU4} on a triangular lattice breaks translational invariance. Hole doping this state in the limit of infinite interactions leads to a state where bosonic holons (red) move freely in an SU(4) symmetric background of fermionic spinons (four flavors represented by different shades of blue). \textbf{b)} SU(4) spin correlations on nearest-neighbor bonds of the doped state are antiferromagnetic. Bond colors are proportional to $\langle \suS{i} \cdot \suS{j} \rangle$. \textbf{c)} The electronic momentum distribution $n_e(\kvec)$ shows a large Fermi surface on top of a constant background for small hole doping $\delta = 1/21$. Presented data are obtained with SU(4)-symmetric matrix product states with a U(1)-equivalent bond dimension of $D=36000$ on infinite 5-leg cylinders (green dashed line in \textbf{b}).}
\label{fig::kinetic_magnetism}
\end{figure}

In this work, we demonstrate that doping mobile charge carriers into a frustrated system with strong interactions provides an alternative route to fractionalized spin excitations. In three dimensions, it was recently argued that a related approach can stabilize an archetypal spin liquid, the resonating valence bond (RVB) state on a pyrochlore lattice~\cite{glittumResonantValenceBond2025}. Here, we investigate a two-dimensional triangular SU(4) model in the limit of strong interactions and low doping. 
We find that the kinetic energy of mobile holes, doped into a Mott insulator, is minimized when the SU(4) spins form a spinon Fermi surface state with fractionalized quasiparticles. We identify the main ingredients to stabilize the spinon Fermi surface: On the one hand, holes suffer from kinetic frustration in the triangular geometry~\cite{haerterKineticAntiferromagnetismTriangular2005, sposettiClassicalAntiferromagnetismKinetically2014, zhuDopedMottInsulators2022a, lebratObservationNagaokaPolarons2024, prichardDirectlyImagingSpin2024}. Not all hopping terms can be minimized simultaneously, in analogy to the geometric frustration for quantum spin liquids. On the other hand, the SU(4) symmetry crucially enhances quantum fluctuations. In fact, for the same model on a triangular lattice with SU(2) symmetry, the hole kinetic energy induces a classical antiferromagnetic order~\cite{haerterKineticAntiferromagnetismTriangular2005, sposettiClassicalAntiferromagnetismKinetically2014}. On the contrary, we show analytically that in the large-$N$ limit of hole-doped SU($N$) systems, the kinetic energy is minimized by a spinon Fermi surface state. Employing unbiased numerical simulations with infinite Matrix Product States (MPS) and variational Monte Carlo (VMC) sampling, we find that the fractionalization persists robustly even for $N=4$.

We propose that the kinetically driven spinon Fermi surface could be realized in moiré heterostructures of transition-metal dichalcogenides (TMDs) and comment on possible detection via the hole spectral function or quantum oscillations.
The high flexibility of these heterostructures not only enables in situ tuning of the electronic filling around the Mott insulator but can also provide additional local degrees of freedom through layer stacking. With two active layers for the electrons, the usual SU(2) spin symmetry can be increased to feature Hubbard models with approximate SU(4) symmetries~\cite{zhangSU4ChiralSpin2021, xuTunableBilayerHubbard2022a, Kuhlenkamp2024, zhangApproximateSU4Spin2023, kimSU4KondoLattice2025}. Ultracold alkaline-earth atoms~\cite{gorshkovTwoorbitalMagnetismUltracold2010, taieObservationAntiferromagneticCorrelations2022, gas-ferrerSpinresolvedMicroscopy$^87$Sr2026} and Rydberg tweezer arrays~\cite{Qiao2025_tJ, qiaoKineticallyinducedBoundStates2025a, martinMeasuringSpectralFunctions2026} offer alternative routes to implement similar models.

\section{Kinetic frustration in SU(4)-symmetric $t$-$J$ models}
Motivated by the experimental realization of multilayer Hubbard models with moiré structures of TMD semiconductors~\cite{xuTunableBilayerHubbard2022a}, we introduce four flavors of electrons corresponding to their spin ($\uparrow$ and $\downarrow$) and layer degrees of freedom (top $t$ and bottom $b$), labeled as:
\begin{equation}
    \ket{1}\equiv\ket{t\uparrow}, \,\ket{2}\equiv\ket{t\downarrow}, \, \ket{3}\equiv\ket{b\uparrow}, \, \ket{4}\equiv\ket{b\downarrow}.
    \label{eq::SU4labels}
\end{equation}
Here, we mainly focus on the case of doping the Mott state of one electron per site $n=1 - \delta$.
For strong interaction, symmetric in the layer and spin degrees of freedom, and $\delta=0$, the system is described by the Kugel-Khomskii model~\cite{kugelJahnTellerEffectMagnetism1982}. Similar to the Heisenberg model for SU(2) degrees of freedom, the Kugel-Khomskii Hamiltonian can be derived from a Schrieffer-Wolff transformation by expanding the underlying Hubbard model for strong interactions $\nobreak{t/U \ll 1}$~\cite{kimSU4KondoLattice2025}:
\begin{align}\label{eq::Heisenberg_SU4}
    H_{J} &= J \sum_{\langle i j \rangle} \left( \suS{i} \cdot \suS{j} + \frac{1}{4}\right)  \\
    &= J \sum_{\langle i j \rangle} \left(2 \mathbf{S}_i \cdot \mathbf{S}_j + \frac{1}{2} \right) \left(2 \mathbf{P}_i \cdot \mathbf{P}_j + \frac{1}{2} \right), \nonumber
\end{align}
on a triangular lattice with nearest neighbors $\langle i j \rangle$, and the superexchange $J=4t^2/U$. Here, we denote the generalized SU(4) spins as $\suS{i}$ and explicitly show the decomposition into spin and layer degrees of freedom, $\mathbf{S}_i$ and $\mathbf{P}_i$, respectively. For a single electron per site, these operators correspond to the fundamental representation of SU(4). To form a fully antisymmetric SU(4) singlet, four sites are needed, $\ket{\psi^\text{asym}_{ijkl}} = \sum_{\mu \nu \sigma \tau} \varepsilon^{\mu \nu \sigma \tau} \ket{\mu_i \nu_j \sigma_k \tau_l} /\sqrt{24} $. Here, Greek letters denote the four different flavors of~Eq.~\eqref{eq::SU4labels} and $\varepsilon^{\mu \nu \sigma \tau}$ is the antisymmetric Levi-Civita symbol.

Numerical studies of that model at filling $n=1$ on the triangular lattice suggest that the ground state spontaneously breaks translational invariance towards a striped~\cite{keselmanEmergentFermiSurface2020, jinUnveilingCriticalStripy2022, zhangVariationalMonteCarlo2024} or plaquette-ordered phase~\cite{pencQuantumPhaseTransition2003, zhangSU4ChiralSpin2021, kimSU4KondoLattice2025, keselmanSU4HeisenbergModel2023}; see Fig.~\ref{fig::kinetic_magnetism}a for a sketch and Appendix~\ref{app::plaquettes} for numerical data.

To introduce holes with a density $\delta$ into the system, we include empty sites $\ket{0}$, resulting in a $t$-$J$ model generalized to SU(4) degrees of freedom~\cite{xuTopologicalSuperconductivityTwisted2018, heGutzwillerApproximationApproach2022}:
\begin{align}
    H &= H_t + H_J \label{eq::tJ_model}, \\
    \mathrm{with}\quad H_t &= -t\sum_{\langle i j\rangle, \alpha}\mathcal{P}_\mathrm{GW}\left(\cd{i\alpha} \co{j\alpha} + \mathrm{h.c.} \right)\mathcal{P}_\mathrm{GW}.\label{eq::hopping_su4}
\end{align}
Here, $\alpha$ describes the four spin flavors, and the Gutzwiller projection $\mathcal{P}_\mathrm{GW}$ only allows singly occupied or empty sites due to the strong onsite repulsion $U$. $H_J$ is given by Eq.~\eqref{eq::Heisenberg_SU4}. The electron operators are related to the SU(4) spins $\suS{i}$ by $\mathcal{S}_i^\mu = \sum_{\alpha, \beta} \cd{i\alpha} \lambda^\mu_{\alpha \beta} \co{i\beta}$, where $\lambda^\mu$ are the 15 generators of SU(4).

In the following, we will focus on the limit of infinite interactions $J\rightarrow 0$. According to Nagaoka's theorem~\cite{Nagaoka1966} and its extensions~\cite{tasakiExtensionNagaokasTheorem1989, whiteDensityMatrixRenormalization2001, newbyFiniteTemperatureKineticFerromagnetism2025, arnoldStrongcouplingFunctionalRenormalization2025, zhangFiniteTemperatureDopantinduced2025}, doping a single hole in this limit leads to a ferromagnet on bipartite lattices. What happens on non-bipartite lattices is, in general, a more challenging question~\cite{haerterKineticAntiferromagnetismTriangular2005, sposettiClassicalAntiferromagnetismKinetically2014, kimExactHoleinducedResonatingvalencebond2023, moreraHightemperatureKineticMagnetism2023, Morera_Attraction,sherifHaerterShastryKineticMagnetism2025} and has also been recently studied experimentally~\cite{dehollain2020nagaoka, xuFrustrationDopinginducedMagnetism2023, prichardDirectlyImagingSpin2024, lebratObservationNagaokaPolarons2024, ciorciaroKineticMagnetismTriangular2023}. 
The complexity of this problem arises from kinetic frustration due to the non-bipartite structure of the lattice. Consider first an SU(2) symmetric system and one hole hopping around a single triangle of a triangular lattice. The hole accumulates a phase factor of~$-1$. As a consequence, to minimize its kinetic energy, the spins must form a singlet, compensating for this minus sign. For a single triangle, we can write the ground state as $\ket{\psi_0} =(\ket{0_1s_{23}} + \ket{0_2 s_{13}} + \ket{0_3s_{12}})/\sqrt{3} $ where $s_{ij}$ describes an SU(2) singlet on sites $i$ and $j$ and $0_k$ denotes the hole on site $k$ \cite{lebratObservationNagaokaPolarons2024,prichardDirectlyImagingSpin2024,martinMeasuringSpectralFunctions2026}.
On specific geometries such as the Husimi cactus~\cite{kimExactHoleinducedResonatingvalencebond2023} or the 3D pyrochlore lattice~\cite{glittumResonantValenceBond2025}, this can be achieved exactly for each triangle, but for the 2D triangular lattice, not all bonds can be simultaneously put into pairwise singlets, therefore, leading to kinetic frustration. For SU(2) symmetric models, this frustration gives rise to the formation of a classical antiferromagnetic state with a three-site sublattice 120$^\circ$ order~\cite{haerterKineticAntiferromagnetismTriangular2005, sposettiClassicalAntiferromagnetismKinetically2014, moreraHightemperatureKineticMagnetism2023, sherifHaerterShastryKineticMagnetism2025}.

For SU(4) symmetry, the hole accumulates a phase factor of $-1$ as well when hopping around a triangle (Fig.~\ref{fig::kinetic_magnetism}a), which can be compensated by an antisymmetric spin wavefunction. Proceeding to the full 2D triangular lattice, the larger number of flavors allows the formation of a fully antisymmetric spin wavefunction for four sites, but beyond that, the system remains kinetically frustrated.
To solve this problem for the SU(4) $t$-$J$ model on the triangular lattice, we therefore numerically search directly for the ground state of Eq.~\eqref{eq::hopping_su4} using unbiased infinite density matrix renormalization group (iDMRG) simulations on various cylinder geometries. Specifically, we focus our analysis on infinitely long cylinders with a circumference of $L_y=5$ and an MPS unit cell size of 21 sites, with spiral boundary conditions (SC5); see Appendix~\ref{app::MPS_geometries} for details. This geometry includes several high-symmetry points in momentum space and allows for competing magnetic orders, such as 120$^\circ$ order or plaquette order of Eq.~\eqref{eq::Heisenberg_SU4} at quarter filling. We consider a finite hole density with one hole per MPS unit cell, e.g., $\delta = 1/21$ for the 21-site unit cell. 

Upon doping holes into the system, our iDMRG simulations variationally find states with antiferromagnetic nearest-neighbor SU(4) spin correlations, indicated by blue bonds $\braket{\suS{i} \cdot \suS{j}} < 0$ in Fig.~\ref{fig::kinetic_magnetism}b. In contrast to the plaquette states that are the ground state of the undoped model Eq.~\eqref{eq::Heisenberg_SU4} at $n=1$, we do not find any signs of translational symmetry breaking in the doped case. Strikingly, in Fig.~\ref{fig::kinetic_magnetism}c, the electronic momentum distribution $n_e(k)$ shows a Fermi sea of occupied states $n=1$ (dark blue) with a Fermi surface that is much larger than naively expected from the low hole concentration. Instead, the Fermi surface corresponds to one-quarter filling of the free particle dispersion on the triangular lattice.
Moreover, for momenta outside the Fermi sea, the electron density does not drop to zero as would be for a Fermi liquid, but instead stays at a finite value of $\sim (1-\delta)$ 
across the full Brillouin zone.

\section{Spinon Fermi surface state}

\begin{figure}[t]
\centering
\includegraphics[trim={0cm 0cm 0cm 0cm},clip,width=\linewidth]{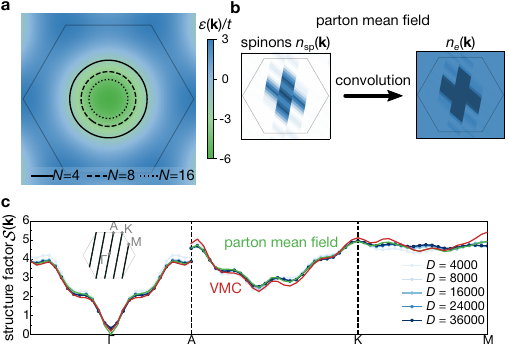}
\caption{\textbf{Spinon Fermi surface.} \textbf{a)} Single-particle dispersion $\varepsilon(\kvec)$ and Fermi surfaces for several fillings $\nu=1/N$. On a triangular lattice, they are approximately circular for sufficiently large $N$. \textbf{b)} A parton mean-field approach predicts that the electronic momentum distribution $n_e(\kvec)$ is given by a convolution of the spinon Fermi surface and the holon distribution, see Eq.~\eqref{eq::convolution}. We show mean-field values of the cylinder geometry studied numerically in Fig.~\ref{fig::kinetic_magnetism} to facilitate their comparison. \textbf{c)} The SU(4) structure factor $\mathcal{S}(\kvec)$ for the same geometry along distinct momentum cuts for a parton mean field (green), variational Monte Carlo (red), and MPS with several bond dimensions up to 36000 (blue).}
\label{fig::partons}
\end{figure}

To understand our numerical findings and identify the origin of the large Fermi surface, we adopt a parton picture for the holes. For general SU($N$) symmetric systems in the large-$N$ limit, parton mean-field solutions can become exact due to a suppression of gauge fluctuations with $1/N$~\cite{wenQuantumFieldTheory2007}. Moreover, for the Kugel-Khomskii model in Eq.~\eqref{eq::Heisenberg_SU4}, it was shown that a self-consistent parton mean-field analysis can accurately capture the plaquette-ordered state~\cite{zhangSU4ChiralSpin2021, kimSU4KondoLattice2025}; see also Appendix~\ref{app::plaquettes}.
To derive the mean-field Hamiltonian of the $t$-$J$ model in the large-$N$ limit~\cite{sachdevLargeLimitSquarelattice1990}, we decompose the fermion creation operators of the underlying model into fermionic spinons~$\fd{i\alpha}$ and bosonic holons~$\bo{i}$:
\begin{align}
    \cd{i\alpha} &= \bo{i} \fd{i\alpha} \\
    \mathcal{S}_i^\mu &= \sum_{\alpha, \beta} \fd{i\alpha} \lambda^\mu_{\alpha \beta} \fo{i\beta}, 
\end{align}
where $\lambda^\mu$ are now the generators of SU($N$).
Moreover, to generalize the SU(4) case of one fermion or hole per site to SU($N$), we impose the onsite constraint:
\begin{equation}\label{eq::su4_constraint}
    \sum_\alpha \fd{i\alpha}\fo{i\alpha} + b^\dagger _i b_i = M.
\end{equation}
Here, $M$ is the maximal number of electrons per site of the corresponding Mott insulator we want to dope. Furthermore, we define the filling factor $\nu=M/N$. When taking the limit $N\rightarrow\infty$, we can either keep $M$ fixed or scale it proportionally to $N$ to maintain a fixed total filling $\nu$. Here, we stick to $M=1$, ensuring that there is exactly one electron or hole per site. This corresponds to the fundamental representation of SU($N$).

Introducing the mean fields $\chi_{ij} = \sum_\alpha \braket{\fd{i\alpha}\fo{j\alpha}}$ and $\zeta_{ij} = \braket{b^\dagger_i b_j}$, we obtain the mean-field Hamiltonian for the kinetic energy in the infinitely strongly interacting limit ($J=0$):

\begin{equation}
    \label{eq::parton_hopping}
    H_t^\mathrm{mf} = -t\sum_{\braket{ij}} \left( \zeta_{ji}\sum_\alpha \fd{i\alpha}\fo{j\alpha} +\chi_{ij} \bd{j} \bo{i} - \zeta_{ji}\chi_{ij} +\mathrm{h.c.} \right).
\end{equation}
In this limit, the spinons directly determine the hopping phases of the holons, and vice versa, the holons give rise to an emergent gauge field, which determines the hopping for the spinons.
To counteract the kinetic frustration associated with hole doping, we search for a self-consistent solution that keeps both the holon and spinon hoppings flux-free, i.e., $\zeta_{ij}>0$, $\chi_{ij}>0$. This minimizes the energy according to the Perron-Frobenius theorem.

The optimal single-particle dispersion is shown in Fig.~\ref{fig::partons}a with a minimum in the center of the Brillouin zone $\varepsilon(\kvec = \Gamma) = -6t$. 
By fixing $M=1$, increasing $N$ leads to a smaller filling per spinon flavor, hence reducing the size of the Fermi surface.
Consider the extreme case, where we have a single hole on a finite lattice and increase $N$ until it reaches the system size $L^2-1$. Then there is only a single spinon per flavor, which can hop around freely, such that its kinetic energy is optimal when $\zeta_{ij}=\mathrm{const.}>0$ for all nearest-neighbor bonds. For an evenly distributed holon at the $\kvec = \Gamma$ point, $\zeta_{ij} = \braket{b^\dagger_i b_j} = 1/L^2$. 
With Eq.~\eqref{eq::parton_hopping}, this also induces a flux-free hopping for the holon, $\chi_{ij} = \sum_\alpha \braket{\fd{i\alpha}\fo{j\alpha}}=1$. The total kinetic energy is:
\begin{equation}
    \label{eq::single_hole_parton_energy}
    E_t^\mathrm{mf} = -t\sum_{\braket{i,j}}(\chi_{ij}\zeta_{ji}+\mathrm{h.c.})= -6t.
\end{equation}
This is the minimal possible value for a single hole on the triangular lattice.

Let us now generalize this argument to the thermodynamic limit with a finite hole density $\delta$. The holons condense at the minimum of the triangular-lattice dispersion at $\kvec = \Gamma$, so $\braket{\bd{i}} = \braket{\bo{i}} = \sqrt{\delta}$. This induces a uniform flux-free hopping for the spinons, which form a Fermi surface. From Fig.~\ref{fig::partons}a, we see that at the considered fillings, the spinon Fermi surface is approximately circular for the triangular lattice. When each spinon flavor has a filling of $(1-\delta)/N$, the absolute value of the Fermi wavevector is given by $k_F = \sqrt{4\pi(1-\delta)/N}$ and we obtain for the nearest-neighbor correlations $\chi_{ij}$:
\begin{align*}
    \chi_{ij} = \sum_\alpha \braket{\fd{i\alpha} \fo{j\alpha}} &= \sum_\alpha  \int \frac{d^2\kvec}{(2\pi)^2} e^{i\kvec(\rvec_i - \rvec_j)} \braket{\fd{\kvec\alpha} \fo{\kvec\alpha}} \\
    &= \frac{N}{(2\pi)^2}\int_0^{k_F} dk \int_0^{2\pi}\,d\phi \,ke^{ik\cos\phi}  = \frac{k_F}{2\pi} J_1(k_F) \\
    &\approx (1-\delta) \quad \mathrm{for}\quad N \gg 4\pi (1-\delta),
\end{align*}
where $J_1(k_F)$ is the Bessel function of the first kind. Therefore, the mean-field energy becomes $E_t^\mathrm{mf} = -t\sum_{\braket{i,j}} (1-\delta) \delta \approx-6tL^2 \delta$; hence, approaching the optimal kinetic energy per hole as well.

\begin{figure}[t]
\centering
\includegraphics[width=\linewidth]{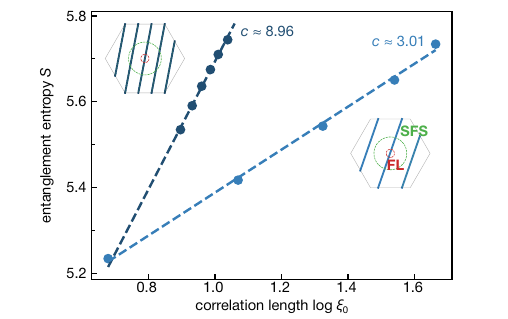}
\caption{\textbf{Central charge of quasi-1D matrix product states.} We fit the central charge from the slope of Eq.~\eqref{eq::central_charge} for infinite MPS simulations of several bond dimensions up to $D=36000$ for different simulated cylinder geometries with spiral boundary conditions SC5 (dark blue) and SC3 (light blue). The insets show the allowed momenta in the hexagonal Brillouin zone (solid lines). The MPS transfer matrix yields the largest correlation lengths in the charge-neutral sector, indicated by the index $\xi_0$. Consistent with a spinon Fermi surface (SFS; green dashed lines), each cut through the Fermi surface contributes an SU(4) fermionic mode with $c=3$ to the total central charge. Contrary, a paramagnetic Fermi liquid (FL) at dopings $\delta =1/21$ ($1/25$) would yield a central charge $c=4$ for both cylinder geometries due to the small Fermi surface (red dashed lines). For better visualization, we added a constant to $S$ for the SC3 cylinder.}
\label{fig::central_charge}
\end{figure}

The parton analysis demonstrates a possible solution for how the doped holes circumvent kinetic frustration by fractionalizing into holons and spinons, both of which are frustration-free. However, the previous considerations required the large-$N$ limit. To determine whether $N=4$ is already sufficiently large, we compare the predictions from the parton mean field with our unbiased iDMRG results. 
First, the spinon-holon decomposition readily explains the observed behavior of the electronic momentum distribution $n_e(\kvec)$. Due to the fractionalization, $n_e(\kvec)$ is given by a convolution:
\begin{equation}
\label{eq::convolution}
    n_e(\kvec) = \frac{1}{L^2} \sum_\qvec n_\mathrm{sp}(\qvec)\left(1+n_h(\kvec-\qvec)\right),
\end{equation}
where the spinons form a Fermi surface $n_\mathrm{sp}(\qvec) = \theta(\qvec < \kvec_F)$ and the holons condense $n_h(\qvec=\Gamma) = L^2\delta$, such that $n_e(\kvec) = 1$ for momenta inside the spinon Fermi surface and $n_e(\kvec) = 1-\delta$ otherwise. Placing the parton mean field on the same cylinder geometry as the MPS yields a similar electron density as in the MPS simulations; compare Fig.~\ref {fig::kinetic_magnetism}b and Fig.~\ref {fig::partons}b.

Next, we check how well observables are captured by the parton mean-field ansatz, and compare the static structure factor to the MPS results:
\begin{equation}
    \label{eq::structure_factor}
    \suS{}(\kvec) = \sum_{i, j} e^{i\kvec(\rvec_i - \rvec_j)} \langle \suS{i} \cdot \suS{j} \rangle.
\end{equation}
For the mean-field ansatz, the parton constraint Eq.~\eqref{eq::su4_constraint} is only fulfilled on average, implying that, for $i=j$, the SU(4) generators do not add up to the correct constant $\langle\suS{i}\rangle_\text{mf}^2 \neq 15/4$. Accordingly, we rescale the mean-field structure factor to compensate for this difference.
As we can see in Fig.~\ref{fig::partons}c, the rescaled results agree quantitatively with the MPS results along the possible momentum cuts in the Brillouin zone, shown in the inset. Furthermore, we explicitly perform a Gutzwiller projection of the spinon Fermi surface state and evaluate the SU(4) expectation values with a Variational Monte Carlo (VMC) simulation. We find that the projection does not significantly alter the structure factor.

The doped states are intrinsically gapless and therefore challenging to represent with MPS, which can only represent finite correlation lengths $\xi$. Utilizing the full SU(4) symmetry, we perform DMRG simulations with an equivalent of a U(1) bond dimension of $D=36000$ and analyze the scaling of the entanglement entropy in these states. For critical states in one dimension, the entanglement entropy scales according to~\cite{calabreseEntanglementEntropyConformal2009, pollmannTheoryFiniteEntanglementScaling2009}:
\begin{equation}
\label{eq::central_charge}
    S = \frac{c}{6} \log\xi.
\end{equation}
Here, $c$ is the central charge of the corresponding conformal field theory. For infinite MPS, we extract the correlation length directly from the largest eigenvalue of the transfer matrix and fit the central charge by simulating several bond dimensions; see Fig.~\ref{fig::central_charge}. Moreover, explicit quantum numbers in the MPS framework provide further information in which charge sector of the transfer matrix the largest eigenvalues can be found.
In two dimensions, gapless states are not necessarily described by a conformal field theory, and therefore, not all gapless states are characterized by a central charge. However, for quasi-one-dimensional cylinder geometries that we consider numerically, we can still extract a meaningful central charge, which provides information about the number of gapless modes~\cite{geraedtsHalffilledLandauLevel2016}.

Crucially, this analysis reveals important differences between the spinon Fermi surface state and a paramagnetic Fermi liquid of the doped holes.
SU($N$)-symmetric critical states in one dimension have a central charge $c=N-1$~\cite{affleckCriticalBehaviourSUn1988}. Hence, for the spinon Fermi surface with $N=4$, we expect a contribution of $c=3$ for each momentum cut through the large spinon Fermi surface. Depending on whether we use the 5-leg MPS cylinders SC5 (dark colors) or the 3-leg MPS cylinders SC3 (light colors), different momentum cuts in the Brillouin zone pass through the Fermi surface; see insets in Fig.~\ref{fig::central_charge}. We observe a large central charge $c\approx9$ for the SC5 geometry, corresponding to three cuts through the spinon Fermi surface, compared to just one momentum cut for the SC3 geometry with $c\approx3$ accordingly. 
In contrast, for a scenario without fractionalization, the holes would form a paramagnetic Fermi liquid. The densities of $\delta=1/21$ and $1/25$ for the two cylinder geometries would result in small Fermi surfaces (red dashed lines in the insets of Fig.~\ref{fig::central_charge} serve as guides to the eye). For free fermions without the constraint in Eq.~\eqref{eq::su4_constraint}, the central charge is $c=1$ per flavor, and therefore, we would expect the single momentum cut to yield a total central charge $c=4$ for both cylinder geometries in the case of a paramagnetic Fermi liquid; see also the case of particle doping with small Fermi surfaces on top of a Nagaoka ferromagnet in Appendix~\ref{app::nagaoka}.

In both geometries, we find the largest correlation lengths in the $\Delta n = 0$ charge sector rather than in $\Delta n = 1$, indicating that spin correlations are more dominant than charge correlations, opposite to the behavior in a normal Fermi liquid. We can understand this from the parton picture as well. Since any charge correlation function will also be a convolution of the holons and spinons, we expect charge correlations to decay more quickly than spin correlations, where only the spinons are involved.

\section{Stability of the spinon Fermi surface}

\begin{figure}[t]
\centering
\includegraphics[trim={0cm 0cm 0cm 0cm},clip,width=\linewidth]{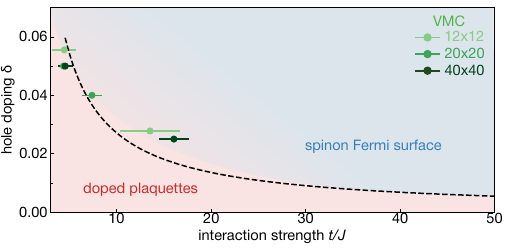}
\caption{\textbf{Stability of the spinon Fermi surface state at finite $t/J$.} From variational Monte Carlo sampling (VMC), we estimate the critical hopping strengths $t_c/J$ by analyzing the competition between kinetic energy, favoring the spinon Fermi surface, and interaction energy, favoring SU(4) plaquettes, for several dopings and system sizes (green color intensity). We estimate the phase diagram in the regime of low dopings and strong interactions. The dashed line represents a fit $t_c/J = \mathrm{const.}/\delta$ to the VMC data. For interactions $t/J\lesssim 1$ (not shown), a full Hubbard description is necessary instead of the $t$-$J$ model to capture other metallic phases.}
\label{fig::stability}
\end{figure}

The infinitely strongly interacting limit of $J\rightarrow0$ is helpful for understanding how minimizing the kinetic energy can induce a spinon Fermi surface. Although interactions in realistic experimental scenarios can be extremely large, they will always be finite. Therefore, we investigate the stability of the spinon Fermi surface for $J>0$. The competing phases are directly deduced from the parton mean-field ansatz.

At quarter filling ($n=1$), it was shown that a mean-field decomposition of Eq.~\eqref{eq::Heisenberg_SU4} can accurately capture the plaquette state as the ground state of the Kugel-Khomskii term~\cite{zhangSU4ChiralSpin2021, kimSU4KondoLattice2025}:
\begin{equation}
    \label{eq::parton_J} 
    H_{J} = -J \sum_{\braket{i,j}}\left( \chi_{ji} \fd{i}\fo{j} +\mathrm{h.c.} -|\chi_{ij}|^2 \right),
\end{equation}
where we assume SU($4$) symmetry and therefore drop the flavor index.
It was pointed out in Ref.~\cite{kimSU4KondoLattice2025} that the plaquette order arises from an instability by considering $\chi_{ij}<0$ in Eq.~\eqref{eq::parton_J}. Starting from a spinon Fermi surface state with the reversed sign, the system features perfectly nested Fermi surfaces with instability vectors $\kvec=\mathrm
M$, which drive the system into a plaquette-ordered state. Therefore, for any finite $J$, we have a competition between the interaction term, which prefers $\chi_{ij}<0$ with translation-symmetry breaking, and the kinetic term Eq.~\eqref{eq::parton_hopping}, which is minimized by the flux-free, uniform state $\chi_{ij}>0$.

To analyze this instability, we employ variational Monte Carlo simulations, which explicitly perform Gutzwiller projections on the Slater determinants of the following mean-field states:
\begin{align}
    \label{eq::H_VMC}
    H_\mathrm{VMC} &= -\sum_{\braket{i j}, \alpha} t_{ij} \fd{i\alpha} \fo{j\alpha} + \mathrm{h.c.} \\
    t_{ij} = t[1+\eta_{ij}h],& \quad \eta_{ij} = 
    \begin{cases}
    +1& \quad \braket{ij} \in \mathrm{plaquettes} \\
    -1& \quad \braket{ij} \notin \mathrm{plaquettes} 
    \end{cases} \nonumber
\end{align}
This Hamiltonian captures the spinon Fermi surface in the limit $h \rightarrow 0$. For any finite $h$, a gap forms at quarter filling, and the plaquette order develops~\cite{keselmanSU4HeisenbergModel2023}. For small dopings, we compute the hole kinetic energy as well as the SU(4) spin-interaction energy for perturbations $h$ around the spinon Fermi surface.

The intuitive picture from above is reproduced. The hole kinetic energy is minimal for $h=0$, i.e., when the spinons form a uniform state with a Fermi surface. In contrast, the interaction energy is lowered by increasing the plaquette term~$h$, as expected from the instability of the Kugel-Khomskii model.
A simple estimate of the transition between the phases is obtained from the energy increase under the quadratic gap opening with $h$, $E_h = E_{h=0} + (at\delta - bJ)h^2$, where we obtain $a$ and $b$ from the VMC data. The spinon Fermi surface state is stable for $(at\delta - bJ)>0$. The estimated critical interactions depend on the doping and allow us to sketch a phase diagram; Fig.~\ref{fig::stability}. At large interactions, even for low hole densities, the kinetic energy is dominant, and a spinon Fermi surface is stabilized. Lowering the interaction strength $t/J$ drives the system into a doped version of the plaquette state to minimize the SU(4) spin interactions. Increasing the hole density from such a state will favor the kinetic energy and therefore, realize a spinon Fermi surface state again. When the interactions become too small, $t/J \lesssim 1$, a full Hubbard model description is required to correctly describe the metallic behavior in that limit.

\section{Implementation and detection in moiré heterostructures}

\begin{figure*}[t]
\centering
\includegraphics[trim={0cm 0cm 0cm 0cm},clip,width=\linewidth]{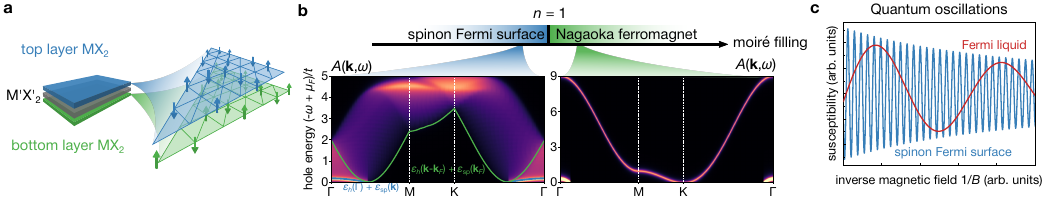}
\caption{\textbf{Experimental realization and signatures.} \textbf{a)} A trilayer heterostructure of TMDs to realize an approximately SU(4)-symmetric Hubbard model. Lattice mismatches between the top and bottom layers with the middle layer generate moiré potentials, leading to strong interactions on a triangular lattice with electrons having spin and layer quantum numbers. \textbf{b)} By tuning the filling around $n=1$, we predict a spinon Fermi surface on the hole-doped site, and a Nagaoka ferromagnet for the electron-doped side. The hole spectral function $A(\kvec, \omega)$ distinguishes these two scenarios. The fractionalization in the spinon Fermi surface leads to a convolution of the spinon dispersion $\varepsilon_\mathrm{sp}(\kvec)$ and the holon dispersion $\varepsilon_h(\kvec)$. The corresponding lines along energy minima are sketched in green and blue, respectively. In the ferromagnet, the spectrum consists of a free dispersion and a contribution from the small Fermi surfaces around the $\Gamma$ point. \textbf{c)} Sketch of quantum oscillations to distinguish between a small Fermi surface of a paramagnetic Fermi liquid with low doping $\delta =1/20$ (red) and a fractionalized state with large Fermi surface (blue).}
\label{fig::tmds}
\end{figure*}

To realize the SU(4) symmetric $t$-$J$ model, Eq.~\eqref{eq::hopping_su4}, experimentally, one needs a triangular lattice with localized SU(4) degrees of freedom and large interactions. Both arise naturally in TMD heterostructures~\cite{zhangSU4ChiralSpin2021, Kuhlenkamp2024, zhangApproximateSU4Spin2023, kimSU4KondoLattice2025}. In a heterostructure, as in Fig.~\ref{fig::tmds}a, lattice mismatches create moiré potentials between the two outer layers, e.g., WS$_2$, and a different material (e.g., MoSe$_2$) in the middle. Alternatively, the moiré potentials can be generated by two twisted hBN layers, which imprint a superlattice on the nearby TMD layers~\cite{kiperConfinedTrionsMottWigner2025a, mhenniEngineeringStrongCorrelations2025}. By appropriate gating, electrons can be tuned continuously in the two active layers. The middle layer stays charge neutral and acts as a tunneling barrier. In the described setup, the doped electrons possess both spin and layer degrees of freedom, creating the flavors for an SU(4) symmetric model.
The moiré potential generates an effective Hubbard model on a triangular lattice~\cite{wuHubbardModelPhysics2018, tangSimulationHubbardModel2020}. Typical interaction strengths can be as large as $U/t \approx 200$~\cite{ciorciaroKineticMagnetismTriangular2023} or equivalently $t/J \approx 50$, which places the $t$-$J$ model well into the strongly interacting limit and within the regime where fractionalization from kinetic frustration can be realized (Fig.~\ref{fig::stability}).
Considering beyond-onsite interactions will lead to additional spin terms in Eq.~\eqref{eq::Heisenberg_SU4}~\cite{zhangSU4ChiralSpin2021, morales-duranNonlocalInteractionsMoire2022, delreFieldControlSymmetrybroken2024} and therefore, a different state than the plaquette state may be realized in the absence of doping. However, upon doping, these terms are suppressed compared to the kinetic energy, provided $t \delta \gg J$.

Flexible tuning of the electron filling in these moiré heterostructures also allows for investigating the case of electron doping, $n = 1 + \delta$. In contrast to doping holes, there is no kinetic frustration, i.e., hopping of the additional particles around a single triangle already yields a phase $+1$. Therefore, the kinetic energy can be simultaneously minimized for all triangles provided the background is a uniform ferromagnet. This is the essence of Nagaoka's theorem~\cite{Nagaoka1966, tasakiExtensionNagaokasTheorem1989}, which holds for particle doping on any connected lattice and for hole doping on bipartite lattices only. Applied to the SU(4) symmetric model at finite doping densities, one flavor forms the ferromagnetic background, while the remaining three flavors form small unpolarized Fermi surfaces; see Appendix~\ref{app::nagaoka} for more details and numerical data. 

We will now discuss concrete experimental signatures that distinguish the fractionalized state obtained upon hole doping from competing states, such as those obtained upon particle doping or the doped plaquettes.
Quantum oscillations in TMDs~\cite{fouttyMagneticHofstadterCascade2025, nguyenQuantumOscillationsDipolar2026} under an external magnetic field could, for example, detect the size of the Fermi surface, thereby revealing a strong asymmetry between hole and particle doping, $n=1 \pm \delta$, and clearly contrasting the spinon Fermi surface from a conventional Fermi liquid such as the doped-plaquettes states. For instance, the De Haas–Van Alphen effect predicts an oscillating behavior of the magnetic susceptibility, where the periodicity in the inverse magnetic field $1/B$ is inversely proportional to the size of the Fermi surface. This behavior is sketched in Fig.~\ref{fig::tmds}c. For a paramagnetic Fermi liquid with a hole density $\delta = 1/20$ away from integer filling, we expect very long oscillations in $1/B$, while for the spinon Fermi surface, all spinons contribute to yield fast oscillations. We estimate that for moiré periodicities of around $a_M \approx10\,$nm, feasible magnetic fields of up to $B=10\,$T \cite{fouttyMagneticHofstadterCascade2025, nguyenQuantumOscillationsDipolar2026} will be sufficient to detect this oscillatory behavior.

Furthermore, single-particle spectral probes have been predicted to reveal fractionalized excitations in quantum spin liquids~\cite{podolskyMottTransitionSpinliquid2009, tangLowenergyBehaviorSpinliquid2013, Kadow2022, Kadow2024, Jin2024, nyhegnProbingQuantumSpin2025}. Due to the splitting of holes into holons and spinons, the spectral function is given as a convolution of these two. The distinct energy scales of holons and spinons enable an identification of the individual contributions~\cite{bohrdtAngleresolvedPhotoemissionSpectroscopy2018, Kadow2022, Kadow2024} or to identify bound states~\cite{bohrdtPartonTheoryARPES2020, moreranavarroExploringKineticallyInduced2024, nyhegnSpinChargeBoundStates2025a}.
As a novel tool for measuring spectral functions in two-dimensional layered materials, the quantum twisting microscope (QTM) has shown promising results for single- and multilayer graphene~\cite {inbarQuantumTwistingMicroscope2023, birkbeckQuantumTwistingMicroscopy2025}. Theoretical proposals show how the QTM can be used to characterize spin-ordered~\cite{Pichler2024} as well as fractionalized states~\cite{periProbingQuantumSpin2024, Pichler2025}.

In Fig.~\ref{fig::tmds}b, we compute the mean-field hole spectral function as measured with the QTM or angle-resolved photoemission spectroscopy (ARPES) in the kinetically induced phases. We plot the energy-resolved spectrum along several momentum cuts in the Brillouin zone, including the high-symmetry points, where we focus on energies below the chemical potential $-\omega +\mu_F > 0$. Here, the spectra are flipped upside down such that the Fermi level is at the bottom.
For the spinon Fermi surface state, the hole spectral function is given as a broad convolution of the spinons and holons:
\begin{align}\label{eq::arpes_parton}
    A(\kvec, \omega) &= \int \mathrm{d}\nu \mathrm{d}\qvec \, A_{\mathrm{sp}}(\kvec-\qvec, \omega-\nu) A_{h}(\qvec, \nu) \nonumber \\
    &= \int \mathrm{d}\qvec \,Z_{\mathrm{sp}}(\kvec-\qvec) Z_{\mathrm{h}}(\qvec) \delta(\omega-\varepsilon_{\mathrm{sp}}(\kvec-\qvec)-\varepsilon_{h}(\qvec)),
\end{align}
with quasiparticle weights $Z_\alpha(\kvec)$. Following the parton construction in Eq.~\eqref{eq::parton_hopping}, we find that spinons have a dispersion $\nobreak{\sim t\zeta_{ij}\sim t\delta}$ and holons are more dispersive $\nobreak{\sim t\chi_{ij}\sim t(1-\delta)}$. Hence, for momenta within the spinon Fermi surface, the lowest branch of the spectrum is given by the spinon dispersion, and the minimum for the holons $\varepsilon_h(\Gamma)+\varepsilon_\mathrm{sp}(\kvec)$ (blue lines). For larger momenta, the holon momentum compensates for the difference to the closest momentum of the spinon Fermi surface (green lines). This analysis is based solely on the mean-field dispersions, and the different spectral weights originate from the density of states of the holons and spinons at each momentum. Possible interactions between holons and spinons beyond the mean-field ansatz will modify the spectrum, but are suppressed with $1/N$ and should qualitatively still yield the same picture. Recent numerical work on the spectral function of a chiral spin liquid has further confirmed that the parton picture captures the main spectral features well~\cite{Kadow2022}.

When considering particle doping, the single-hole spectral features are very distinct: the spectral function has two main contributions. The excitation can either move in the fully occupied band and thereby resolve the triangular-lattice dispersion. Alternatively, a dynamical hole can be removed from the small Fermi sea at low momenta. This reveals the small Fermi surfaces around the $\Gamma$ point at low energies, but with higher intensities due to the added contributions of the three remaining flavors.

\section{Outlook}

The quest for fractionalized phases of matter---where microscopic degrees of freedom reorganize into emergent quasiparticles carrying fractional quantum numbers---is a central challenge in strongly correlated systems. Most routes to fractionalization build on geometric frustration and topological order. Here, we take a different perspective and present a concrete example of a two-dimensional system, where kinetic frustration and strong interactions stabilize a spinon Fermi surface upon doping holes into an SU(4) symmetric quantum magnet.

From an experimental perspective, moiré platforms offer an appealing setting to study this phenomenon because of their high tunability in filling and interaction scales. A key challenge for the future will be to identify signatures of fractionalization in experimentally accessible probes, beyond the most direct (but often demanding) momentum-resolved spectral measurements. For instance, nonlinear pump-probe spectroscopy~\cite{choiTheoryTwoDimensionalNonlinear2020, mcginleySignaturesFractionalStatistics2024, evrardAcStarkSpectroscopy2025} can provide promising insights into low-lying excitations.
The detection of spin liquid states in the Mott insulating regime using noise spectroscopy with nitrogen-vacancy (NV) centers in diamond has been theoretically proposed ~\cite{chatterjeeDiagnosingPhasesMagnetic2019, khooProbingQuantumNoise2022}. We expect that this technique can also reveal a state with a kinetically induced spinon Fermi surface.

Advances in the preparation and cooling of alkaline-earth atoms in optical lattices have enabled the realization of correlated Mott insulators and the observation of antiferromagnetic correlations with general SU($N$) degrees of freedom~\cite{ Taie2012_SuN_Mott, Hofrichter_SuN_Mott, Tusi2022_SuN_Mott, taieObservationAntiferromagneticCorrelations2022}. Based on our analytic arguments in the large-$N$ limit, we expect larger values $N>4$ to stabilize the spinon Fermi surface state even further. Combined with spin-resolved quantum gas microscopy~\cite{gas-ferrerSpinresolvedMicroscopy$^87$Sr2026}, these systems provide a promising platform for probing the exotic spinon Fermi surface states. Two-photon Raman spectroscopy offers a direct route to measuring the dynamical spin structure factor~\cite{Prichard_Raman_2025}. At zero doping, one expects a gap $\Delta$, with no spectral weight below it and a sharp onset at frequencies $\omega \sim \Delta$. This feature corresponds to breaking a local singlet into a triplet excitation, a magnon-like quasiparticle (triplon), which should manifest as a sharp, dispersing mode. Upon doping, however, the formation of a spinon Fermi surface leads to a closing of the gap. The sharp spectral features are then expected to broaden significantly, with a linewidth set by the size of the spinon Fermi surface, reflecting the fractionalized excitations. Momentum- and energy-resolved single-particle spectral functions are experimentally accessible as well, for example, by radio-frequency transfer into a non-interacting state~\cite{Brown2019}.

On the theoretical level, it will be crucial to assess the stability of fractionalization from kinetic frustration against perturbations to the idealized model studied in this work, including weak breaking of SU(4) symmetry and longer-ranged interactions. Looking ahead, an important question is the \emph{generality} of this phenomenon: what are the minimal requirements for kinetic frustration to favor parton fractionalization rather than conventional symmetry breaking? For example, a related phenomenon has been recently put forward for stabilizing RVB states on a pyrochlore lattice~\cite{glittumResonantValenceBond2025}. Addressing these questions may help delineate a broader class of systems and materials where kinetic frustration can be harnessed for realizing fractionalized quantum matter.

Interestingly, other systems, which are not inherently SU(4) symmetric, may realize a similar behavior~\cite{wuHIDDENSYMMETRYQUANTUM2006}. Consider, for instance, a spin $S=3/2$ system with arbitrary spin interactions:
\begin{align}
    \label{eq::tJ_s3_2}
    H_{S=3/2} &= -t\sum_{\langle i j\rangle, \alpha}\mathcal{P}_\mathrm{GW}\left(\cd{i\alpha} \co{j\alpha} + \mathrm{h.c.} \right)\mathcal{P}_\mathrm{GW} \nonumber \\
    &+ \sum_{\langle ij \rangle} J_1 (\mathbf{S}_i\cdot \mathbf{S}_j) + J_2 (\mathbf{S}_i\cdot \mathbf{S}_j)^2 + J_3 (\mathbf{S}_i \cdot \mathbf{S}_j)^3 + \dots,
\end{align}
where $\dots$ denote additional spin terms. While the spin part of the Hamiltonian is drastically different from the SU(4) case Eq.~\eqref{eq::Heisenberg_SU4}, the hole hopping also describes the exchange of four flavors on a triangular lattice. Therefore, in the limit of infinite interactions $J_i \rightarrow 0$, the model obtains an enlarged SU(4) symmetry and is the same as we explore in this work, hinting at a broad applicability of the proposed mechanism. 

We focus on the limit of very low doping to study the effects of kinetic magnetism on the SU(4) spin. In this regime, we find no indication of superconductivity, see Appendix~\ref{app::superconductor}. However, upon increasing hole doping, these systems may realize superconducting states~\cite{heGutzwillerApproximationApproach2022, xuTopologicalSuperconductivityTwisted2018, classenCompetingPhasesInteracting2019}. 
Understanding the underlying mechanism of such superconducting states will be an interesting future direction.

\section*{Acknowledgments}
WK thanks Sunghoon Kim for insightful discussions about the nature of the plaquette states. WK and MK acknowledge support from the Deutsche Forschungsgemeinschaft (DFG, German Research Foundation) under Germany’s Excellence Strategy--EXC--2111--390814868, TRR 360 – 492547816, and DFG grants No. KN1254/1-2, KN1254/2-1, the European Union (grant agreement No. 101169765), as well as the Munich Quantum Valley (MQV), which is supported by the Bavarian state government with funds from the Hightech Agenda Bayern Plus. IM and ED acknowledge support from the SNSF project 200021-212899; the SNSF Sinergia grant CRSII--222792; NCCR SPIN, a National Centre of Competence in Research, funded by the Swiss National Science Foundation (grant number 225153); the Swiss State Secretariat for Education, Research and Innovation (contract number UeM019-1).

\section*{Data availability} All data, data analysis, and code for generating the data are available on Zenodo~\cite{zenodo}.

\appendix
\section{Plaquette states for the Mott insulator at quarter filling}
\label{app::plaquettes}

\begin{figure}[t]
\centering
\includegraphics[trim={0cm 0cm 0cm 0cm},clip,width=\linewidth]{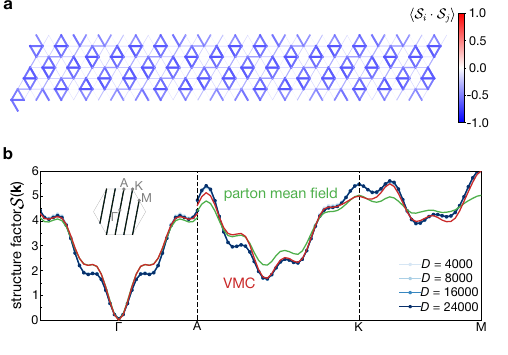}
\caption{\textbf{MPS results at $n=1$.} \textbf{a)} Nearest-neighbor SU(4) spin correlations $\langle \suS{i} \cdot \suS{j} \rangle$ reveal a translational symmetry-breaking ground state of 4-site plaquettes. The presented data are for an MPS with bond dimension $D=24000$ and a MPS unit cell consisting of eight SC5 unit cells. Bond thickness and color are proportional to $\nobreak{\langle \suS{i} \cdot \suS{j} \rangle}$ \textbf{b)} The static structure factor $\suS{\kvec}$ computed with MPS of bond dimensions up to $D=24000$ is similar to the one from a rescaled mean-field ansatz (green) and VMC calculations (red).}
\label{fig::plaquettes}
\end{figure}

The ground state of the Kugel-Khomskii model Eq.~\eqref{eq::Heisenberg_SU4} at $n=1$ has been studied in other contexts before~\cite{pencQuantumPhaseTransition2003, keselmanEmergentFermiSurface2020, jinUnveilingCriticalStripy2022, zhangVariationalMonteCarlo2024, zhangSU4ChiralSpin2021, kimSU4KondoLattice2025, keselmanSU4HeisenbergModel2023}. For small cylinder sizes, several DMRG studies have yielded different results. While for YC$3$ cylinders with a circumference of $L_y=3$, no symmetry-breaking pattern was found, YC4 cylinders with $L_y = 4$ show emergent critical stripes~\cite{keselmanEmergentFermiSurface2020, jinUnveilingCriticalStripy2022}. On even larger cylinders of $L_y =8, 10$, the stripes turn into plaquettes~\cite{zhangSU4ChiralSpin2021}, similar to the ones we observe on the spiral SC5 geometry with $L_y =5$.

These results can be understood from a mean-field instability analysis, introduced in Ref.~\cite{kimSU4KondoLattice2025}. Consider the spinon mean-field Hamiltonian of the spin interaction term:
\begin{equation*}
    H_{\tilde{J}} = -\tilde{J} \sum_{\braket{i,j}}\left( \chi_{ji} \fd{i}\fo{j} +\mathrm{h.c.} -|\chi_{ij}|^2 \right).
\end{equation*}
For uniform $|\chi_{ij}|$ at quarter filling, these states form a Fermi surface. For $\chi_{ij} > 0$, this Fermi surface will be circular around the center of the Brillouin zone as in Fig.~\ref{fig::partons}a. Conversely, for $\chi_{ij} < 0$, the Fermi surfaces will be split around the $\kvec=\mathrm{K}$ points and are perfectly nested, such that bond-density-wave instabilities occur at the $\kvec=\mathrm{M}$ points of the Brillouin zone. In the 2D thermodynamic limit, these instabilities lead to plaquette order~\cite{kimSU4KondoLattice2025}. However, performing a similar analysis for the finite-circumference cylinders reveals that there is no such instability for the YC3 geometry and only one of the $M$ points has an instability for the YC4 cylinder, explaining the observed DMRG results.
We also note that recent large-scale variational Monte Carlo simulations indicate that within the plaquettes, the SU(4) symmetry is spontaneously broken~\cite{keselmanSU4HeisenbergModel2023}.

We compare the parton approach to the infinite DMRG simulations for the SC5 geometry. We find that these cylinders also exhibit instabilities towards the translation symmetry-broken plaquette state. We can directly observe these plaquettes by computing $\braket{\suS{i} \cdot \suS{j}}$ on nearest-neighbor bonds, see Fig.~\ref{fig::plaquettes}a. In order to get an overall SU(4) singlet with the correct instability vectors, we need to enlarge the MPS unit cell by repeating the 21-site unit cell eight times along the $\rvec_x$ direction.
Moreover, we compare the static structure factor $\suS{}(\kvec)$ evaluated from the MPS to that from a parton mean-field ansatz, Eq.~\eqref{eq::H_VMC}, and its projected version from variational Monte Carlo (VMC). Here, we determine the strength $h$ of the plaquette term by minimizing the SU(4) spin interaction energy on the cylinder geometry in the VMC simulations and find the minimal energy for $h_\mathrm{min} \approx 0.25 t$.
The results of the parton mean field are rescaled to fit the SU(4) algebra, $\langle\suS{i}\rangle^2 = 15/4$, which is not enforced when the constraint $\sum_\alpha \fd{i\alpha} \fo{i\alpha}=1$ only holds on average.

\section{Details on MPS simulations and cylinder geometries}
\label{app::MPS_geometries}

\begin{figure}[t]
\centering
\includegraphics[trim={0cm 0cm 0cm 0cm},clip,width=\linewidth]{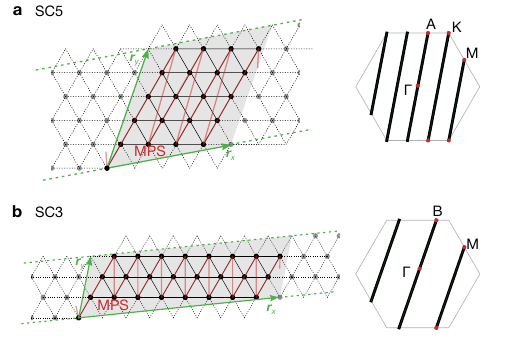}
\caption{\textbf{MPS on spiral cylinder geometries.} \textbf{a)} Left: SC5 cylinder geometry with a 21-site unit cell with periodic boundary conditions along $r_y$ and infinite boundaries along $r_x$, indicated by the dashed lines. The order of MPS sites snaking through the cylinder is shown in red. Right: Brillouin zone with allowed momenta for the SC5 cylinder, including the high symmetry points $\Gamma, M,$ and $K$. \textbf{b)} The SC3 cylinder geometry with a 25-site unit cell has a smaller cylinder circumference (left), and therefore, fewer cuts through the Brillouin zone in momentum space (right).}
\label{fig::MPS_geometries}
\end{figure}

Simulating the SU(4) symmetric $t$-$J$ model with matrix product states (MPS) on the triangular lattice is a challenging task due to the large local Hilbert space dimension $d=5$ (4 flavors + empty/doubly occupied sites) and the inherently gapless nature of the states upon doping. To perform the infinite density matrix renormalization group (iDMRG) studies, we use the TeNPy~\cite{Hauschild2024} and MPSKit~\cite{vandammeMPSKit2025} libraries. 
In TeNPy, we allow states that break the full SU(4) symmetry. This is particularly efficient for computing the ground state in the case of particle doping, where a fully polarized Nagaoka ferromagnet is realized. Instead we conserve the independent U(1) charges $n, S^z, P^z,$ and $S^zP^z$. Fermionic exchange statistics are realized by Jordan-Wigner strings.

For the hole-doping scenario, we find that the U(1) symmetries are not sufficient to faithfully represent the ground state. Therefore, we utilize the MPSKit library, which features non-Abelian symmetries. We conserve the total particle number $n$, explicit fermion parity, and the full SU(4) symmetry of the models Eqs.~\eqref{eq::Heisenberg_SU4} and ~\eqref{eq::hopping_su4}. We keep up to 2000 SU(4) multiplets, corresponding to a U(1) bond dimension of up to 36000. 

\begin{figure}[t]
\centering
\includegraphics[trim={0cm 0cm 0cm 0cm},clip,width=\linewidth]{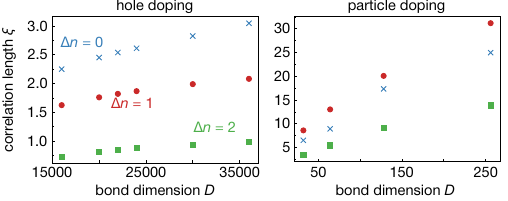}
\caption{\textbf{MPS correlation lengths in different charge sectors.} We compute the MPS correlation lengths via the transfer matrix of the SC5 geometry in different charge sectors $\Delta n = 0, 1, 2$ for states with increasing bond dimensions for hole doping and particle doping.}
\label{fig::correlation_lengths}
\end{figure}

Compared to previous DMRG studies on the triangular lattice~\cite{keselmanEmergentFermiSurface2020, heGutzwillerApproximationApproach2022}, we choose different cylinder geometries with spiral boundary conditions as depicted in Fig.~\ref{fig::MPS_geometries}, where we impose periodic boundary conditions along $\rvec_y$ and repeat the MPS unit cell (red) infinitely often along $\rvec_x$. The unit cell sizes of 21 sites and 25 sites are chosen to simulate low but finite doping of $\delta = 1/21, 1/25$, respectively, and allow for the surrounding SU(4) spins to form an overall singlet state, while still keeping the unit cell sizes modest, required for reasonable iDMRG update times. The SC5 geometry with 21 sites per unit cell, primarily studied in this work, has a circumference of $L_y=5$ with an effective MPS distance between nearest-neighbor sites that is lower than in the more often used YC4 cylinders. Additionally, the allowed momenta in the Brillouin zone include the $\kvec= \mathrm{K}$ and M points. Therefore, the MPS can form plaquette states at $n=1$ (compare with Appendix \ref{app::plaquettes}) or could host the 120$^\circ$ order observed for doped SU(2) models. In fact, the same unit cell with 21 sites was used by Haerter and Shastry for exact diagonalization calculations~\cite{haerterKineticAntiferromagnetismTriangular2005}.

\section{Absence of superconductivity}
\label{app::superconductor}

In the main text, we use the central charge to identify the nature of the doped states. This is computed as a direct relation between the entanglement entropy and the correlation length of (quasi) one-dimensional critical systems, Eq.~\eqref{eq::central_charge}. While the entanglement entropy is directly obtained from the MPS singular values at each bond, the correlation length can be computed for each symmetry sector with the transfer matrix, defined for infinite MPS~\cite{hauschildEfficientNumericalSimulations2018}. We show the correlation lengths $\xi$ for different charge sectors $\Delta n = 0, 1, 2$ in Fig.~\ref{fig::correlation_lengths}. For both hole doping and particle doping, $\xi$ increases with bond dimension in all charge sectors. This indicates that all the corresponding states are gapless in each charge sector. In particular, we do not see any signs of superconducting pairs, since $\xi_{\Delta n =1} > \xi_{\Delta n =2}$ and both increase with bond dimension.

We note that in our scenario, condensation of the bosons does not directly imply superconductivity via the Anderson-Higgs mechanism~\cite{sachdevLargeLimitSquarelattice1990, wenEffectiveTheoryBreaking1989}. To see this, we couple the holons $b$ to the internal gauge field $a_\mu$ and to an externally applied field $A_\mu$, so the Lagrangian is $\mathcal{L}_b \propto |(\partial_\mu - a_\mu - A_\mu)b|^2$. The spinons $f_\alpha$ do not carry charge and therefore only see the internal gauge field coupled with $-a_\mu$, $\nobreak{\mathcal{L}_f \propto \sum_\alpha|(\partial_\mu +a_\mu)f_\alpha|^2}$. We can associate two gauge symmetries with $a_\mu$ and $A_\mu$. When the bosons condense, they can break one of these gauge symmetries, but the other one will still be unbroken and coupled to the gapless spinons. Hence, the electromagnetic gauge field is not necessarily gapped out. It will be an interesting direction for future investigation to determine whether larger doping or additional spin interactions can lead to a superconducting state.

\section{Nagaoka ferromagnetism}
\label{app::nagaoka}

In the main text, we study the effect of hole doping, $n=1-\delta$. However, the high tunability in moiré heterostructures or ultracold atoms similarly allows for doping additional electrons, $n=1+\delta$. Numerically, we study the effect of particle doping with the same Hamiltonian as in Eq.~\eqref{eq::hopping_su4}, but $\mathcal{P}_\mathrm{GW}$ projects onto states with exactly one or two electrons per site.

\begin{figure}[b]
\centering
\includegraphics[trim={0cm 0cm 0cm 0cm},clip,width=\linewidth]{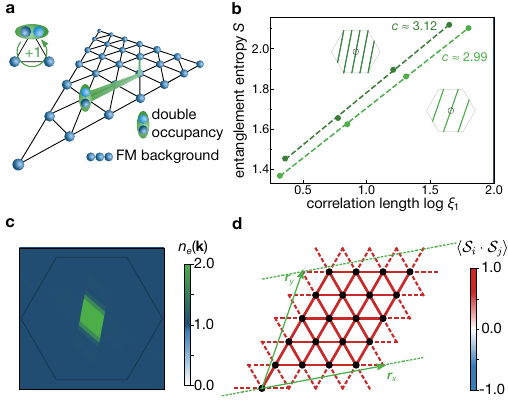}
\caption{\textbf{Nagaoka ferromagnetism for particle doping.} \textbf{a)} Additional particles doped above $n=1$ can move around freely in a fully polarized ferromagnetic background. \textbf{b)} The entanglement entropy scaling from infinite MPS simulations reveals a central charge $c\approx3$ for both the SC5 and SC3 cylinder geometries (insets). This is consistent with a small Fermi surface, which \textbf{c)} is also directly detected by the electronic momentum distribution $n_e(\kvec)$. \textbf{d)} The nearest-neighbor correlations $\braket{\suS{i} \cdot \suS{j}}$ are ferromagnetic, in agreement with Nagaoka's theorem.}
\label{fig::nagaoka_FM}
\end{figure}

For a single particle above quarter filling, the ground state is captured by Nagaoka's theorem at infinite interaction and we hence expect a ferromagnetic ordering. Intuitively, we can understand this by looking at a single triangle. A particle hopping around this triangle yields a phase of $+1$, Fig.~\ref{fig::nagaoka_FM}a. Hence, all hopping paths interfere constructively, and the kinetic energy of the additional particles is minimized by aligning all background SU(4) spins into a fully symmetric wave function.

We use infinite DMRG simulations to investigate the behavior at finite electron densities $n=1+1/21$ on the SC5 cylinder geometry; see Fig.~\ref{fig::nagaoka_FM}. We find a Nagaoka ferromagnet, with $\braket{\suS{i} \cdot \suS{j}}> 0$ on all bonds. The electron density $n_e(\kvec) = \sum_\alpha \cd{\kvec\alpha} \co{\kvec\alpha}$ reveals a small Fermi sea (green) $\sim \delta$ on top of the uniform background with $n=1$ (dark blue). 
The additional particles form three equal, small Fermi seas in the center of the Brillouin zone. 
The central charge is obtained from Eq.~\eqref{eq::central_charge}. We observe the same central charge $c\approx3$ for both cylinder sizes. Since the additional particles can move around freely in a ferromagnetic background, we expect a free fermion central charge of $c=1$ per flavor for each momentum cut through the Fermi sea. Therefore, one flavor forms the ferromagnetic background while the other three contribute $c=1$ to the central charge. Moreover, in contrast to the hole-doping case, we find the largest eigenvalue in the $\Delta n = 1$ charge sector of the transfer matrix, as expected from a Fermi liquid; Fig.~\ref{fig::correlation_lengths}.

\bibliography{references}

\end{document}